\documentclass[twocolumn,showpacs,preprintnumbers,unsortedaddress,%
floatfix,amsmath,prl]{revtex4}
\usepackage{bm}% bold math
\usepackage{graphicx}%
\usepackage{calc}%
\usepackage{amsfonts}%
\usepackage{times}
\begin{document}
\title{
Why Kohn-Sham and Hartree-Fock orbitals are very close to each other.\\
Shell structure of exchange potential in atoms.}
\author{ M. Cinal}
\affiliation{Institute of Physical Chemistry of the Polish Academy
of Sciences, ul. Kasprzaka 44/52, 01--224 Warszawa, Poland}
\date{\today}
\begin{abstract}

%The exact  local exchange potential of
%the density functional theory
%is found to be very well represented, in closed-shell atoms, by a piecewise  function built of the potentials obtained
%with the non-local Fock exchange operator for  the  Hartree-Fock (HF) orbitals.
%This explains the outstanding proximity of Kohn-Sham and HF orbitals and
% the high quality of the Krieger-\-Li-Iafrate
% and  localized HF
%approximations to the exchange potential.

It is found that, in closed-$l$-subshell atoms, the exact  local exchange potential of the density functional theory 
is very well represented, within the region of every atomic shell, 
by  each  of the  suitably shifted potentials obtained with the non-local Fock exchange operator 
acting on the Hartree-Fock (HF) orbitals that belong to this shell.
This explains the outstanding proximity of Kohn-Sham and HF orbitals and
the high quality of the Krieger-\-Li-Iafrate  and  localized HF (or, equivalently, common-energy-denominator) 
approximations to the exchange potential.

%It is found
%that, in closed-shell atoms, the exact  local exchange potential of
%the density functional theory
%is very well represented by a piecewise  function built of the potentials obtained
%with the non-local Fock exchange operator for  the occupied Hartree-Fock (HF) orbitals.
%This explains the outstanding proximity of Kohn-Sham and HF orbitals as well as
% the high quality of the Krieger-\-Li-Iafrate
% and  localized HF (common-energy-denominator)
%approximations to the exact  exchange potential.

\end{abstract}
\maketitle
Representing the quantum state of a many-electron system in terms of
one-electron orbitals is simple and theoretically attractive
approach. Such description is realized in the Hartree-Fock (HF)
method \cite{johnson07}, as well as in the Kohn-Sham (KS) scheme of
the density-functional theory (DFT)  \cite{FN03} which is
an efficient and robust tool,
now routinely applied in the calculations of electronic properties of
molecules and condensed-matters structures, even very large and complex.
Though the
KS scheme is formally accurate, the one-body KS potential contains
the exchange-correlation (xc) potential $v_{\text{xc}}$, whose dependence on the
electron density remains unknown. It is usually treated within the
local density or generalized-gradient
approximations (LDA, GGA),
despite the well-known shortcomings of the LDA and GGA xc potentials
(especially the self-interaction errors). Some of these
deficiencies are removed when the exact form (in terms of the KS occupied orbitals)
is used for the exchange part $E_{\text{x}}$ of the xc energy.
The exact
exchange potential  $v_{\text{x}}$ is then found from $E_{\text{x}}$ by means of the integral
equation resulting from the optimized-effective-potential (OEP)
approach [\onlinecite{KLI92}(a),\onlinecite{EV93,KK08}] or by using the
recently developed method based on the differential equations for
the orbital shifts \cite{KP03,CH07}. The exact potential $v_{\text{x}}$  is %exchange potential is
free from self-interaction and it has correct asymptotic dependence
($-1/r$ for finite systems) at large distances $r$  from the system;
thus, unlike the HF, LDA or GGA potentials, it produces correct unoccupied states.
In the DFT, the approximation, in which the exchange is
included exactly but the correlation energy and potential are
neglected, is known as the exchange-only KS scheme --- it is applied
in the present investigation.
The full potential $v_{\text{xc}}$ can also be found by means of the OEP approach
when the DFT total energy includes, besides the exact $E_{\text{x}}$,
the correlation energy $E_{\text{c}}$ depending on all (occupied
and unoccupied) KS orbitals. % and orbital energies.
This makes such computation tedious, to a level undesirable in the DFT,
since it involves calculating $E_{\text{c}}$ with  the quantum-chemistry
methods, like the M{\o}ller-Plesset many-body perturbation approach.

Defined to yield the true electron density,
the KS one-electron orbitals  have no other direct physical meaning
since they formally refer to a {\em  fictitious} system of {\em non-interacting}
electrons. However, it is a common practice to use these orbitals in
calculations of various electronic properties; in doing so the
$N$-electron ground-state wave function $\Psi_0$ of the physical
(interacting) system is approximated with the single determinant
built of the KS orbitals. This approximate approach is justified by
(usually) sufficient accuracy of the calculated quantities, which is
close to, or often better than, that of the HF results. It seems that the
success of the DFT calculations would not be possible if the  KS
determinant, though being formally non-physical, was not close to
the HF determinant which, outside the DFT,   is routinely
used to approximate the wave function $\Psi_0$  of
the {\em real} system. Therefore, understanding this proximity is
certainly very important for the fundamentals of the DFT.

Previous calculations \cite{ZP92_93,CES94}
have shown that,
not only the whole KS and HF determinants, but also
the individual occupied KS and HF orbitals
$\phi_{a\sigma}({\bf r})$ and $\phi_{a\sigma}^{\text{HF}}({\bf r})$
in atoms are so close to each other that
they are virtually indistinguishable
(here the orbitals, dependent on the electron position $\bf r$
and the spin $\sigma=\downarrow,\uparrow$,
are numbered with index $a=1,\ldots,N_{\sigma}$; $N_{\downarrow}+N_{\uparrow}=N$).
This proximity is particularly remarkable for the exchange-only KS orbitals.
It is surprising in view  of the
obvious difference between the exchange operators
in the KS and HF one-electron hamiltonians (see below) and the fact that
the corresponding KS and HF atomic orbital energies,
$\epsilon_{a\sigma}$ and $\epsilon_{a\sigma}^{\text{HF}}$,
differ substantially, up to several hartrees for core orbitals in atoms
like Ar, Cu \cite{EV93,KK08}.
This apparent contradiction has not yet been resolved;
in Ref. \cite{IL03}  it is suggested
that the KS and HF determinants are close to each other
``since the kinetic energy is much
greater than the magnitude of the exchange energy".
The present paper investigates the proximity of the KS and HF orbitals and
reveals that
it results from the specific properties of
the HF  {\em orbital } exchange potentials \cite{KLI92}
\begin{equation}
v_{\text{x}a\sigma}^{\text{HF}}({\bf r}) \equiv
\left[ \hat{v}_{\text{x}\sigma}^{\text{F}}\phi_{a\sigma}^{\text{HF}}({\bf r}) \right]
/\phi_{a\sigma}^{\text{HF}}({\bf r}) \; .
\label{eq-vxa}
\end{equation}
obtained with the non-local Fock exchange operator
$\hat{v}_{\text{x}\sigma}^{\text{F}}$
built of the HF orbitals.
 Simultaneously, it becomes
clear why,  in atoms, the DFT KS exact exchange potential
$v_{\text{x}\sigma}(r)$ has the characteristic structure of a
piecewise-like function where each part spans over the region of an
atomic shell and it has distinctively different slope
$dv_{\text{x}\sigma}(r)/dr$  in consecutive shells \cite{leeuwen96}.
These shell constituents of $v_{\text{x}\sigma}(r)$ are found in the
present Letter to be very well represented, within the region
of each atomic shell, by the suitably shifted potentials
$v_{\text{x}a\sigma}^{\text{HF}}(r)$, Eq. (\ref{eq-vxa}),  for the HF orbitals
$\phi_{a\sigma}^{\text{HF}}({\bf r})$ that belong to this shell.
Thus, it is shown how these {\em orbital-specific} HF potentials,
which describe the exchange in the physical system, are almost
directly mapped onto the KS local exchange potential
$v_{\text{x}\sigma}(r)$ which is {\em common} to all electrons with spin $\sigma$ in
the KS non-interacting system. The revealed properties
of $v_{\text{x}a\sigma}^{\text{HF}}({\bf r})$ are also found to stand
behind the high quality of the Krieger-\-Li-Iafrate(KLI)
\cite{KLI92} approximation to the exact potential
$v_{\text{x}\sigma}({\bf r})$, as well as, of the localized HF
(LHF)\cite{DSG01} approximation, equivalent to the
common-energy-denominator approximation (CEDA) \cite{GB01}. Note
that the KLI potential can be derived by assuming that the
exchange-only KS and HF orbitals are identical while the LHF (CEDA)
potential is found when the  KS and HF determinants are assumed
to be equal.

The HF one-electron spin-orbitals $\phi_{a\sigma}^{\text{HF}}(\bf
r)$ are obtained by minimizing the mean value $\langle \Psi |\hat
H|\Psi \rangle$ where $\hat H$ is the Hamiltonian of the
$N$-electron interacting system and $\Psi $ belongs to the subspace
of normalized $N$-electron wave functions that are single
determinants built of one-electron  orbitals. Similar  minimization
is carried out in the (exchange-only) OEP method, but there are two additional
constraints ($\sigma=\uparrow,\downarrow$) that for every trial
determinant all $N_{\sigma}$ constituent spin-orbitals
$\phi_{a\sigma}({\bf r})$ satisfy the KS  equation with some local KS
potential $v_{\text{s}\sigma}({\bf r})$. The  minimizing potential
$v_{\text{s}\sigma}({\bf r})=v_{\text{s}\sigma}^{\text{OEP}}({\bf
r})$, yields, after subtracting from it the external
$v_{\text{ext}}({\bf r})$ and electrostatic $v_{\text{es}}({\bf r})$
terms, the exact  exchange potential $v_{\text{x}\sigma}({\bf r})=
v_{\text{x}\sigma}^{\text{OEP}}({\bf r})$. Thus, the KS equation,
satisfied by the corresponding (OEP) orbitals $\phi_{a\sigma}({\bf
r})$ and their energies
 $\epsilon_{a\sigma}$, takes the form
\begin{equation}
    \Bigl(-\frac{1}{2}\bm{\nabla}^2 +
    v_{\text{ext}}+v_{\text{es}}[n_{\text{tot}}]+
    v_{\text{x}\sigma}\Bigr)
    \phi_{a\sigma} =
\epsilon_{a\sigma}\phi_{a\sigma} \, .
\label{eq-ks}
\end{equation}
The total electron density
$n_{\text{tot}}({\bf r})=n_{\uparrow}({\bf r})+n_{\downarrow}({\bf r})$,
which enters
$v_{\text{es}}[n_{\text{tot}}]({\bf r})=\int d\, {\bf r'}\,
n_{\text{tot}}({\bf r'})/|{\bf r'-r}|$,
is the sum
of the
spin-projected densities
$n_{\sigma}({\bf r})=\sum_{a=1}^{N_{\sigma}} |\phi_{a\sigma}({\bf r})|^2$.
In the HF equations, satisfied by the orbitals
$\phi_{a\sigma}^{\text{HF}}(\bf r)$ and energies
$\epsilon_{a\sigma}^{\text{HF}}$,
the multiplicative local exchange potential
$v_{\text{x}\sigma}({\bf r})$   is replaced with
the non-local Fock exchange integral operator
$\hat{v}_{\text{x}\sigma}^{\text{F}}({\bf r})$,
built of $\{ \phi_{a\sigma}^{\text{HF}} \}_{a=1}^{N_{\sigma}}$,
while the potential $v_{\text{es}}({\bf r})$ is found
for the HF total electron density
defined in a similar way as  $n_{\text{tot}}({\bf r})$.
Real KS and HF orbitals are used throughout this paper,  dependence on
 $\sigma $ is suppressed hereafter.
Both
the KS and HF equations need to be solved selfconsistently.

The exact exchange potential
$v_{\text{x}}=v_{\text{x}}^{\text{OEP}}$ satisfies the OEP equation \cite{KP03}
\begin{equation}
\sum_{a=1}^N \phi_a({\bf r})\delta\phi_a({\bf r})=0
\; ,\;\;  \forall{\bf r}\;,
\label{eq-oep}
\end{equation}
which results from the OEP minimization
and depends on $v_{\text{x}}$ through the orbital shifts (OS) $\delta \phi_a({\bf r})$.
Each OS fulfills   the equation \cite{KP03,CH07}
(below $\phi_a$, $\epsilon_a$ are the solutions of Eq. (\ref{eq-ks}))
\begin{equation}
\left( \hat{h}_{\text{s}}-\epsilon_a \right)\delta \phi_a({\bf r}) =W_a^{\perp}({\bf r})
\label{diff-eq-dphi}
\end{equation}
and it is subject to the constraint $\langle \phi_a|\delta\phi_a\rangle =0$.
The equation (\ref{diff-eq-dphi}) includes
the   KS Hamiltonian $ \hat{h}_{\text{s}}$, present in Eq. (\ref{eq-ks}), and
the term (defined using the sign convention of Refs. \cite{KP03, CH07})
\begin{equation}
W_a^{\perp}({\bf r})=
\left[\hat{v}_{\text{x}}^{\text{F}}({\bf r})
+D_{aa}-v_{\text{x}}({\bf r}) \right] \phi_a({\bf r}) \, .
\label{w-perp}
\end{equation}
Here, we have
$\hat{v}_{\text{x}}^{\text{F}}=
\hat{v}_{\text{x}}^{\text{F}}\left[\{ \phi_b \}\right]$,
$D_{aa}= \langle \phi_a |v_{\text{x}}-\hat{v}_{\text{x}}^{\text{F}}| \phi_a\rangle$.

The OS $\delta\phi_a$ and the constant $D_{aa}$ give, within the
perturbation theory, the first-order approximations   to the
differences $-(\tilde{\phi}_a^{\,\text{HF}}-\phi_a)$ and
$-(\tilde{\epsilon}_{a}^{\,\text{HF}}-\epsilon_{a})$. Here, the
orbitals $\tilde{\phi}_a^{\,\text{HF}}$ and the corresponding
energies $\tilde{\epsilon}_a^{\,\text{HF}}$, are the solutions of
the HF-like equation which is the same as Eq. (\ref{eq-ks}) except
for $v_{\text{x}}$ replaced by $\hat{v}_{\text{x}}^{\text{F}}$ built
of the KS orbitals $\{ \phi_b \}$. The corresponding perturbation is
then equal to
$\delta\hat{h}_{\text{s}}=\hat{v}_{\text{x}}^{\text{F}}-v_{\text{x}}$
so that the first-order correction to $\phi_a({\bf r})$, i.e.,
$-\delta \phi_a ({\bf r})= -\sum_{t=1,t \neq a}^{\infty} c_{ta}\,\phi_t({\bf r})$,
$c_{ta}=\langle\phi_t |\delta\hat{h}_{\text{s}}|\phi_a \rangle/(\epsilon_t-\epsilon_a)$,
satisfies Eq. (\ref{diff-eq-dphi}) and the constraint indeed.
 Obviously, the solutions $\tilde{\phi}_a^{\,\text{HF}}$,  $\tilde{\epsilon}_a^{\,\text{HF}}$
are not identical to the self-consistent
HF orbitals $\phi_a^{\text{HF}}$ and orbital energies $\epsilon_a^{\text{HF}}$.
However,
the relations $\| \Delta \phi_a - (-\delta\phi_a) \| < 0.13 \| \Delta \phi_a \|$
(where $\| \phi\|^2=\int d {\bf r} \,|\phi({\bf r})|^2$),
$| \Delta \epsilon_a - (-\delta\epsilon_a) | < 0.003 | \Delta \epsilon_a |$
\cite{note-diff-en}
found for Be and Ar,
validate the representations of
$\Delta \phi_a \equiv \phi_a^{\text{HF}}-\phi_a$ by $-\delta\phi_a$
and
$\Delta \epsilon_a \equiv \epsilon_a^{\text{HF}}-\epsilon_a$ by $\delta\epsilon_a=-D_{aa}$
used in the presented argument;
the above inequalities are obtained for $\phi_a$, $\epsilon_a$,
$\delta\phi_a$ calculated as in %with the method of
Ref. \cite{CH07}, and
$\phi_a^{\text{HF}}$ (expanded in the Slater-type-orbital basis), $\epsilon_a^{\text{HF}}$
taken from Ref. \cite{bunge93}.

The part of
$W_a({\bf r}) \equiv \delta \hat{h}_{\text{s}}({\bf r})\phi_a({\bf r})$
parallel to the orbital $\phi_a$ is $W_a^{||}({\bf r})=-D_{aa}\phi_a({\bf r})$
and it sets the energy shift $\delta \epsilon_a$.
The part $W_a^{\perp}({\bf r})=W_a({\bf r})-W_a^{||}({\bf r})$,
perpendicular to $\phi_a$,
sets  the OS $\delta \phi_a({\bf r})$, Eq. (\ref{diff-eq-dphi}).
Thus, the KS and HF orbitals, $\phi_a({\bf r})$, $\phi_a^{\text{HF}}({\bf r})$,
can be close to each other, even if the orbital energies
$\epsilon_{a}$, $\epsilon_{a}^{\text{HF}}$,
differ significantly, provided  the term $W_a^{\perp}({\bf r})$ is sufficiently small.
Note that the orbitals remain unchanged when a (possibly orbital-dependent) constant
is added to the Hamiltonian in the KS or HF  equations.

For closed-$l$-subshell atom,
the non-local (integral) Fock exchange operator,
acting on an atomic orbital
$\phi_a({\bf r})=r^{-1}\chi_{nl}(r) Y_{lm}(\theta,\varphi)$ ($a\equiv nlm$),
yields
$\hat{v}_{\text{x}}^{\text{F}}({\bf r})\phi_a({\bf r}) =
r^{-1} F_{\text{x};nl}(r) \, Y_{lm}(\theta,\phi)$
where  $Y_{lm}(\theta,\varphi)$ is the spherical harmonic;
hereafter,
$n$, $l$, $m$  denote the principal,  orbital, and magnetic quantum numbers.
The factor $F_{\text{x};nl}(r)$
is defined  \cite{johnson07} with
the exchange integrals  and determines the
$v_{\text{x}a}({\bf r})=v_{\text{x};nl}(r)=
F_{\text{x};nl}(r)/\chi_{nl}(r)$.
The OS
$\delta\phi_a({\bf r})=r^{-1}\delta\chi_{nl}(r) Y_{lm}(\theta,\varphi)$
depends on  $W_a^{\perp}({\bf r})=r^{-1}W_{nl}^{\perp\text{;rad}}(r) Y_{lm}(\theta,\varphi)$
through its radial part
\begin{equation}
\label{wrad-perp}
W_{nl}^{\perp ;\text{rad}}(r)=F_{\text{x};nl}(r)+\left[D_{nl;nl}-v_{\text{x}}(r)\right]\chi_{nl}(r)
\end{equation}
entering the equation for $\delta\chi_{nl}(r)$ found from Eq. (\ref{diff-eq-dphi}).

\begin{figure}
\includegraphics*[width=\columnwidth]{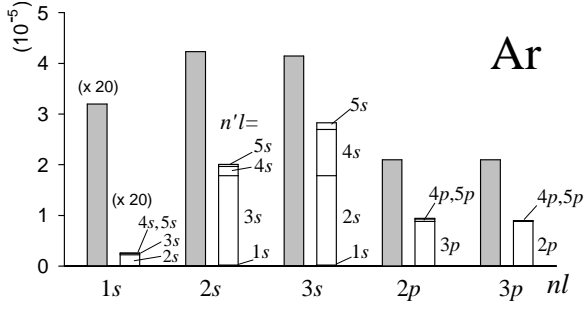}\caption{The OS norm square  $\|\delta \phi_a\|^2$ (grey bars)
and the contributions $c_{n'l;nl}^2$ (stacked bars) to it from bound states
$\phi_{n'lm}$, for the occupied states $\phi_{nlm}$ in the Ar atom;
the $1s$ bars are magnified by the factor 20.
}
\label{fig-dchi-dist-Ar}
\end{figure}

Neither the OEP minimization nor
the resulting OEP equation (\ref{eq-oep}) imply readily
that the orbital shifts $\delta \phi_a({\bf r})$ are small.
However,  the {\em occupied} KS and HF orbitals differ so minutely from each other
that, for atoms,  the OEP total energy %$E$
is only several mhartrees higher than the HF energy $E_{\text{HF}}$
\cite{EV93,KK08,GE95}.
The proximity of individual HF and KS orbitals can be quantified
with the norms $\|\delta \phi_a\|$ which are indeed very small,
in comparison with
$\|\phi_a\|=\|\phi_a^{\text{HF}}\|=1$.
Calculating the  OS $\delta\phi_a$  with the method of Ref.
\cite{CH07}, we  obtain $\|\delta \phi_{1s}\|=0.00669 $,  $\|\delta \phi_{2s}\|=0.00630$ for Be and
$\|\delta \phi_a\|<0.007$ for
each occupied orbital in the Ar atom.
The partition $\|\delta \phi_{nlm}\|^2=\sum_{n' \neq n}^{\infty} c_{n'l;nl}^2$,
plotted for Ar in Fig. \ref{fig-dchi-dist-Ar},
shows that, among the  KS {\em bound} orbitals $\phi_{n'lm}$,
the dominating contributions $c_{n'l;nl}^2$ to the $nlm$ OS
come from
the  $n'lm$ orbitals, $n'=n-1,n+1$, i.e., from
the neighboring electronic shells;
e.g., for  $\delta\phi_{3s}$ in Ar, the largest terms
$c_{n'l;nl}^2$ are found for the $n'l=2s$ (occupied) and $n'l=4s$ (unoccupied) orbitals.
But,  there remains a large  part
of $\|\delta \phi_{nlm}\|^2$
which cannot be attributed to higher unoccupied bound
states $\phi_{n'lm}$ since
the corresponding $c_{n'l;nl}^2$ terms vanish rapidly;
see  Fig. \ref{fig-dchi-dist-Ar}.
This unaccounted part comes from {\em continuum} KS states
($\epsilon_{n'l}>0$).
These results confirm that the assumption
$\delta\phi_{a}=0$ and
$\delta \phi_a - \sum_{t \neq a}^{\text{occ}}c_{ta} \phi_t =0$,
used to derive the KLI and LHF (CEDA) approximation,  respectively,
are very accurate but not exact.

\begin{figure}
\includegraphics*[width=\columnwidth]{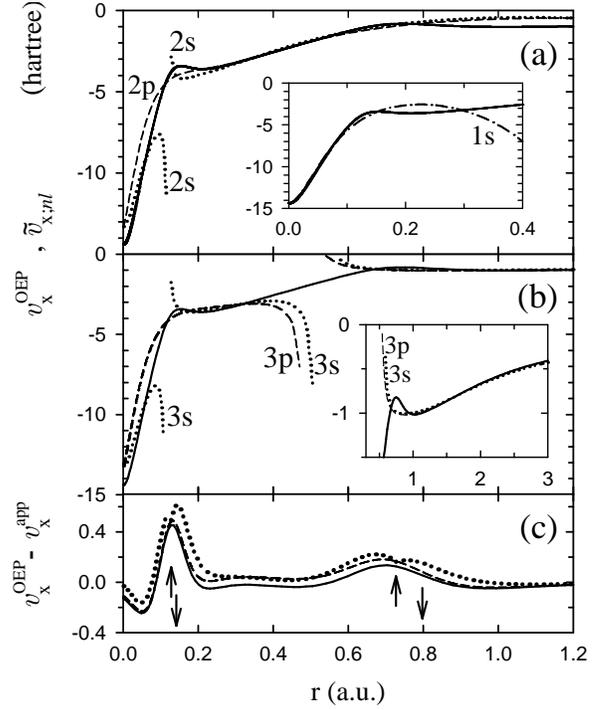}
\caption{ (a,b) The potentials $v_{\text{x}}^{\text{OEP}}$ (solid
line), $\tilde{v}_{\text{x;}nl}$ (dashed and dotted lines) in the Ar
atom; the HF potentials $\tilde{v}_{\text{x;}nl}^{\text{HF}}$ follow  $\tilde{v}_{\text{x;}nl}$
(within the figure resolution);
(c) The differences
$v_{\text{x}}^{\text{OEP}}-v_{\text{x}}^{\text{pw}}$ (dotted line),
$v_{\text{x}}^{\text{OEP}}-v_{\text{x}}^{\text{KLI-HF}}$ (dashed
line),
$v_{\text{x}}^{\text{OEP}}-v_{\text{x}}^{\text{KLI-OEP}}=v_{\text{x}}^{\text{OS}}$
(solid line); the up and down arrows mark the points
$r_{n,n+1}^{\text{HF}}$ and $r_{n}^{\text{min}}$, respectively.
 }
\label{fig-Ar_vx_vxa}
\end{figure}

The norms $\|\delta \phi_{nlm}\|$ have such low values
because the terms $W_{nl}^{\perp;\text{rad}}(r)$
are sufficiently small {\em for all} $r$. %; cf. Fig. \ref{fig-Ar_vx_vxa} (b) $<??>$.
This, combined with the relation
\begin{equation}
%\tilde{v}_{\text{x;}nl}(r)=
v_{\text{x;}nl}(r)+D_{nl;nl}=
v_{\text{x}}(r)+
\frac{W_{nl}^{\perp\text{;rad}}(r)}{\chi_{nl}(r)}\; ,
\label{vxa-tilde}
\end{equation}
found with Eq. (\ref{wrad-perp}), implies that each {\em shifted} KS-OEP
 {\em orbital} exchange potential
\begin{equation}
\tilde{v}_{\text{x;}nl}(r)\equiv v_{\text{x;}nl}(r)+D_{nl;nl}
\end{equation}
is very close to the exact exchange potential
$v_{\text{x}}(r)=v_{\text{x}}^{\text{OEP}}(r)$ within the
$r$-interval $(r_{n-1,n},r_{n,n+1})\equiv S_n$ where the
denominators in the r.h.s. of Eq. (\ref{vxa-tilde}), i.e., the
orbitals $\chi_{nl}(r)$ from the $n$-th atomic shell
($K,L,M,\ldots$), have largest magnitudes; the shell border points
$r_{n,n+1}$ for $n=0,1,\ldots,n_{\text{occ}}$ (the
respective HF points $r_{n,n+1}^{\text{HF}}$, defined precisely
below, can be used) are near the positions $r_{n}^{\text{min}}$ where the radial
electron density $\rho(r)$ has local minima.
The potentials $\tilde{v}_{\text{x;}nl}(r)$ are also
close to  $v_{\text{x}}^{\text{OEP}}(r)$ in the central parts of shells $S_{n'}$,  $n'<n$,
though not so tightly as for the shell $S_n$. %; Fig. \ref{fig-Ar_vx_vxa}.
The proximity of the potentials, evident in Fig.
\ref{fig-Ar_vx_vxa}(a,b) for the Ar atom, holds as well  for other
closed-$l$-subshell atoms.
It
is disturbed  in  the vicinity of the nodes of $\chi_{nl}(r)$,
where the potential $\tilde{v}_{\text{x;}nl}(r)$ diverges while
the  term
$W_{nl}^{\perp\text{;rad}}(r)$
is finite and small.
The potential  $\tilde{v}_{\text{x;}nl}(r)$ also differs significantly from
 $v_{\text{x}}^{\text{OEP}}(r)$ within the shells
$S_{n'}$, $n'>n$, where both the functions $\chi_{nl}(r)$, $W_{nl}^{\perp\text{;rad}}(r)$
decay exponentially.

Since the KS orbitals $\phi_a^{\text{OEP}}=\phi_a [v_{\text{x}}^{\text{OEP}}]$ found for the
exact exchange potential $v_{\text{x}}^{\text{OEP}}$
are very close to   $\phi_a^{\text{HF}}$, the terms
$F_{\text{x;}nl}(r)$, $v_{\text{x;}nl}(r)$,
and $D_{nl,nl}[v_{\text{x}}]$ (for any $v_{\text{x}}$) %, and $\tilde{v}_{\text{x;}nl}(r)$
obtained with $\{\phi_a^{\text{OEP}}\}$
are virtually indistinguishable from
the quantities $F_{\text{x;}nl}^{\text{HF}}(r)$, $v_{\text{x;}nl}^{\text{HF}}(r)$,
$D_{nl,nl}^{\text{HF}}[v_{\text{x}}]$  %,  $\tilde{v}_{\text{x;}nl}^{\text{HF}}(r)$
calculated with
the HF orbitals $\{\phi_a^{\text{HF}}\}$. % and $v_{\text{x}}=v_{\text{x}}^{\text{OEP}}$.
Thus,
the terms
$W_{nl}^{\perp;\text{rad}}[v_{\text{x}},\{\phi_a^{\text{HF}}\}](r)$ are very close to
$W_{nl}^{\perp;\text{rad}}[v_{\text{x}},\{\phi_a^{\text{OEP}}\}](r)$.
%so that
As a result,
they are small for
$v_{\text{x}}=v_{\text{x}}^{\text{OEP}}$, and, also, by continuity, for any potential
%$v_{\text{x}}\approx v_{\text{x}}^{\text{OEP}}$
$v_{\text{x}}$ close to $v_{\text{x}}^{\text{OEP}}$.
Such a class ${\cal V}_0$ \cite{note-class-V0} of potentials  $v_{\text{x}}$ that yield
small $W_{nl}^{\perp;\text{rad}}[v_{\text{x}},\{\phi_a^{\text{HF}}\}] $ %=
and, consequently, lead to
the KS orbitals $\phi_a [v_{\text{x}}]$ almost identical to $\phi_a^{\text{HF}}$
\cite{note-HF-per-theo},
exists provided it is possible, for given $v_{\text{x}}$, to find constants $C_{nl}$ that make terms
$U_{nl}(r) \equiv  F_{\text{x;}nl}^{\text{HF}}+C_{nl}\chi_{nl}^{\text{HF}}-v_{\text{x}}\chi_{nl}^{\text{HF}}$
small {\em for all} $r$.
Indeed, we {\em then} obtain $D_{nl;nl}^{\text{HF}} \approx C_{nl}$ so that the term
$W_{nl}^{\perp;\text{rad}}[v_{\text{x}},\{\phi_a^{\text{HF}}\}]\approx U_{nl}$ is small.
Low magnitude of  $U_{nl}(r)$ implies that, {\em within} each occupied shell $S_n$,
the  shifted HF potentials
$\tilde{v}_{\text{x;}nl}^{\text{HF}}(r)=v_{\text{x;}nl}^{\text{HF}}(r)+C_{nl}$
($l\in {\cal L}_n \equiv \{0,\ldots,l_{\text{max}}^{(n)}\}$)
lie very close to $v_{\text{x}}(r)$, and, as a result, they
almost {\em coincide with  each other},
\begin{equation}
\label{eq-hf-rel-vxnl}
\tilde{v}_{\text{x;}nl}^{\text{HF}}(r) \approx \tilde{v}_{\text{x;}nl'}^{\text{HF}}(r) \; ,  \;\;
l, l' \in {\cal L}_n \; , r\in S_n \; ,
\end{equation}
(similarly, as the OEP potentials $\tilde{v}_{\text{x;}nl}(r)$ do; Fig.
\ref{fig-Ar_vx_vxa}(a,b)). A generalization of Eq.
(\ref{eq-hf-rel-vxnl}) is found when, in the expression for
$U_{nl}(r)$, the potential $v_{\text{x}}(r)$ is replaced
by $\tilde{v}_{\text{x;}n'l'}^{\text{HF}}(r)$ for $r \in S_{n'}$ and
the smallness of $U_{nl}(r)$ is accounted for. The generalized relation reads
\begin{equation}
\label{eq-hf-rel}
F_{\text{x;}nl}^{\text{HF}}+C_{nl}\chi_{nl}^{\text{HF}} \approx
\left(v_{\text{x;}n'l'}^{\text{HF}}+C_{n'l'} \right) \chi_{nl}^{\text{HF}}\, ,\;\;r\in S_{n'}
\end{equation}
and it is satisfied for suitable set of constants $\{C_{nl}\}$ for
all indices $(nl)$, $(n'l')$ corresponding to the occupied HF
orbitals. This is an {\em intrinsic} property of the HF orbitals
(and the Fock operator), since it is {\em not} implied by the DFT or
the definition of  $v_{\text{x}}^{\text{OEP}}$, though it is
revealed here by inspecting the results for
$v_{\text{x}}=v_{\text{x}}^{\text{OEP}}$. The total energy
$E[v_{\text{x}}] \equiv  \langle
\Psi[v_{\text{x}}]|\hat{H}|\Psi[v_{\text{x}}]\rangle$,  for any
$v_{\text{x}} \in {\cal V}_0$, is very close to $E_{\text{HF}}$ (due
to $\phi_a[v_{\text{x}}]\approx \phi_a^{\text{HF}}$), and, as a
result, also to $E[v_{\text{x}}^{\text{OEP}}]$ since the potential
$v_{\text{x}}^{\text{OEP}}$ minimizes the functional
$E[v_{\text{x}}] > E_{\text{HF}}$. This can explain why the exact
exchange potential $v_{\text{x}}^{\text{OEP}}$ belongs to the class
${\cal V}_0$ and, as a result, it gives the KS orbitals $\phi_a$
very close to $\phi_a^{\text{HF}}$.

Assuming that the constants $C_{nl}$ satisfying the relation  (\ref{eq-hf-rel}) are known,
we can construct a continuous {\em piecewise} potential
\begin{equation}
\label{vx-pw}
v_{\text{x}}^{\text{pw}}(r) = \sum_{n}^{\text{occ}}\theta_n^{\text{HF}}(r) v_{\text{x}}^{(n)}(r)
\end{equation}
formed from the HF {\em shell} exchange potentials
\begin{equation}
\label{vx-shell}
 v_{\text{x}}^{(n)}(r) \equiv
 \sum_{l\in{\cal L}_n}
 \left[v_{\text{x;}nl}^{\text{HF}}(r)+C_{nl} \right]  \rho_{nl}^{\text{HF}}(r)/ \rho_n^{\text{HF}}(r)\, ,
\end{equation}
each applied in its shell region $S_n$. The points
$r_{n,n+1}^{\text{HF}}$ defining the shell borders
are the solutions of the continuity equation
%are found from the continuity condition
$v_{\text{x}}^{(n)}(r)=v_{\text{x}}^{(n+1)}(r)$ for
$n=1,2,\ldots,n_{\text{occ}}-1$; $r_{0,1}^{\text{HF}}=0$. We denote
$\rho_{nl}^{\text{HF}}=4\pi (2l+1)(\chi_{nl}^{\text{HF}})^2$,
$\rho_n^{\text{HF}}= \sum_{l\in{\cal L}_n} \rho_{nl}^{\text{HF}}$,
$\rho^{\text{HF}}= \sum_{n}^{\text{occ}} \rho_{n}^{\text{HF}}$,
$\theta_n^{\text{HF}}(r)=\theta(r-r_{n-1,n}^{\text{HF}})\theta(r_{n,n+1}^{\text{HF}}-r)$.
Each shell potential $v_{\text{x}}^{(n)}(r)$ is very close to the
almost coinciding potentials $\tilde{v}_{\text{x;}nl}^{\text{HF}}(r)
\equiv v_{\text{x;}nl}^{\text{HF}}(r)+C_{nl}$, $l \in {\cal L}_n$,
for $r\in S_n$. Thus, the relation (\ref{eq-hf-rel})
with $\tilde{v}_{\text{x;}n'l'}^{\text{HF}}$ replaced by
$v_{\text{x}}^{(n')}$ means that the terms $U_{nl}(r)$  are small
for  $v_{\text{x}}=v_{\text{x}}^{\text{pw}}$. This also holds for
%the Krieger-\-Li-Iafrate(KLI)-{\em like}
the KLI-{\em like}
\cite{KLI92} potential
\begin{equation}
\label{vx-kli}
\breve{v}_{\text{x}}^{\text{KLI-HF}}(r) \equiv
%\sum_{nl}^{\text{occ}}  u_{\text{x;}nl}  \rho_{nl}^{\text{HF}}/ \,\rho^{\text{HF}}=
\sum_{nl}^{\text{occ}} \left(F_{\text{x;}nl}^{\text{HF}}+C_{nl}\chi_{nl}^{\text{HF}}\right)
 \chi_{nl}^{\text{HF}}/ \,\rho^{\text{HF}} \; ,
\end{equation}
since, due to Eq. (\ref{eq-hf-rel}), it is very close to
$\tilde{v}_{\text{x;}n'l'}^{\text{HF}}(r)  \approx
v_{\text{x}}^{(n')}(r)$ for $r \in S_{n'}$, and, thus,
to $v_{\text{x}}^{\text{pw}}$ for all $r$. Therefore, both potentials
$v_{\text{x}}^{\text{pw}}$ and
$\breve{v}_{\text{x}}^{\text{KLI-HF}}$ belong to $ {\cal V}_0$. In
particular,
%it holds for
it can be shown to hold for
$v_{\text{x}}^{\text{KLI-HF}}=\breve{v}_{\text{x}}^{\text{KLI-HF}}[\{C_{nl}^{\text{KLI-HF}}\}]$
where the constants
%are the solution of the equation
are found from their self-consistency condition
$C_{nl}^{\text{KLI-HF}}=D_{nl;nl}[v{_\text{x}}^{\text{KLI-HF}}, %[\{C_{n'l'}^{\text{KLI-HF}}\}],
\{\chi_{n'l'}^{\text{HF}}\} ]$ given in Ref. \cite{KLI92}.
%Indeed, these constants  minimize
%the function $f[\{C_{nl}\}]=\sum_{nl}^{\text{occ}}
% \| W_{nl}^{\perp;\text{rad}}[\breve{v}_{\text{x}}^{\text{KLI-HF}}[\{C_{n'l'}\}]] \|^2$
%(cf. Ref. \cite{CH07}), and, as a result, they should yield small
%$U_{nl}=W_{nl}^{\perp;\text{rad}}$ since
%there exist constants  $\{C_{nl}\}$ that give small values of $U_{nl}[v_{\text{x}}^{\text{KLI-HF}}]$;
%these are the constants that satisfy Eq. %relation
%(\ref{eq-hf-rel}).

The potential $v_{\text{x}}^{\text{pw}}(r)$, built of the HF shell
potentials $v_{\text{x}}^{(n)}(r)$ with
$C_{nl}=C_{nl}^{\text{KLI-HF}}$, is found (see Fig.
\ref{fig-Ar_vx_vxa}(c)) to be a very good approximation to the exact
exchange potential   $v_{\text{x}}^{\text{OEP}}(r)$. The quality of
its approximation is almost the same as of the potentials
$v_{\text{x}}^{\text{KLI-HF}}$ and $v_{\text{x}}^{\text{KLI-OEP}}$.
The latter is defined similarly as
$\breve{v}_{\text{x}}^{\text{KLI-HF}}$ but with the HF orbitals
$\{\chi_{nl}^{\text{HF}}\}$ replaced with the OEP ones
$\{\chi_{nl}^{\text{OEP}}\}$ and with
$C_{nl}=D_{nl;nl}[v_{\text{x}}^{\text{OEP}},\{\chi_{nl}^{\text{OEP}}\}]$.
The potential  $v_{\text{x}}^{\text{KLI-OEP}}$ is the dominant part
of the exact exchange potential
$v_{\text{x}}^{\text{OEP}}=v_{\text{x}}^{\text{KLI-OEP}}+v_{\text{x}}^{\text{OS}}$,
where the minor part $v_{\text{x}}^{\text{OS}}$, depends linearly on
$\{\delta\phi_a\}$ \cite{KP03,CH07}. It is this OS term
$v_{\text{x}}^{\text{OS}}(r)$  that produces the characteristic
bumps of $v_{\text{x}}^{\text{OEP}}(r)$ at the shell borders; cf.
Fig. \ref{fig-Ar_vx_vxa}(c).
%; $v_{\text{x}}^{\text{KLI-OEP}}$ is a smooth, bump-free function.

In summary,
we find that when, for each HF orbital,
a suitably chosen (orbital-specific)
constant is added to the Fock exchange operator in the HF equation,
the electrons occupying different
HF orbitals are subject to very similar local exchange potentials
(and hence they move in very similar total potentials) within the atomic regions
where the orbital radial probability densities are substantial.
This proximity is particularly tight
for the exchange potentials of the orbitals that belong to the same shell
and  it holds in the region of this shell.
Thus, each HF orbital is only very slightly disturbed when the
(shifted) orbital exchange potential is replaced in the HF equation
with a KS exchange  potential (common to all orbitals) that,
within each shell, lies very close to this orbital potential. As a
result, in each shell, the KS exact exchange potential is very well
approximated with the shifted HF orbital exchange
potentials of this shell, and, even better, with their weighted average
-- the shell exchange potential, Eq. (\ref{vx-shell}).

Discussions with A. Holas are gratefully acknowledged.


\begin{thebibliography}{99}
\bibitem{note-diff-en}
The latter is not valid  for $a=\text{HOMO}$, since $D_{aa}=0$,
but  $\Delta\epsilon_{a}=-4.13\cdot 10^{-5}\;\text{a.u.}$ $ (-2.645\cdot 10^{-4}\;\text{a.u.})$ for Be (Ar).
\bibitem{note-class-V0}
The class ${\cal V}_0={\cal V}_0(\delta)$ can be defined precisely
with the condition $\| \phi_a [v_{\text{x}}] -\phi_a^{\text{HF}}\|\leq \delta$
($a=1,\ldots,N$) where $\delta \ll 1$.
\end{thebibliography}
\end{document}